\documentclass[aps,physrev,reprint,showpacs,longbibliography]{revtex4-2}
\usepackage[english]{babel}
\usepackage{amsmath,amssymb,bbm,graphicx,color,comment}
\usepackage[bookmarks=true,colorlinks,citecolor=blue,urlcolor=blue]{hyperref}
\usepackage{dsfont}
\usepackage{soul}

\usepackage{bbm}
\usepackage{float}

\usepackage{dcolumn}   
\usepackage{bm}        
\usepackage{filecontents}
\usepackage{lineno}

\usepackage{mathtools}
\usepackage[usenames,dvipsnames]{xcolor} 

\usepackage[export]{adjustbox}
\usepackage{adjustbox}

\usepackage{braket}
\usepackage{tikz}
\usepackage{bbold}

 \renewcommand{\v}[1]{\ensuremath{\mathbf{#1}}}

\let\baraccent=\= \renewcommand{\=}[1]{\stackrel{#1}{=}}
\newcommand{\eq}[1]{Eq.\thinspace(\ref{#1})}
\newcommand{\eqs}[2]{Eqs.\thinspace(\ref{#1},\ref{#2})}

\newcommand{\fig}[1]{Fig.\thinspace{}\ref{#1}}
\newcommand{\sect}[1]{Sec.\thinspace{}\ref{#1}}
\newcommand{\sects}[2]{Secs.\thinspace{}\ref{#1}-\ref{#2}}
\newcommand{\fc}[1]{({#1})}
\newcommand{\figc}[2]{Fig.\thinspace{}\ref{#1}\thinspace{}\fc{#2}}

\usepackage[normalem]{ulem}

\newcommand{\neel}{\mathbb{Z}_2}
\newcommand{\neelp}{\mathbb{Z}_2^{'}}
\newcommand{\Hpxp}{H_{\text{PXP}}}
\newcommand{\Hpyp}{H_{\text{PYP}}}
\newcommand{\Hpzp}{H_{\text{PZP}}}
\newcommand{\Hpxyp}{H_{\text{PXYP}}}
\begin{document}

\title{
Floquet engineering of interactions and entanglement \\ in periodically driven Rydberg chains
}

\author{Nazl\i \ U\u{g}ur K\"oyl\"uo\u{g}lu$^{1,2}$}
\author{Nishad Maskara$^{1}$}
\author{Johannes Feldmeier$^{1}$}
\author{Mikhail D. Lukin$^{1}$}
\affiliation{$^1$Department of Physics, Harvard University, Cambridge, MA 02138, USA \\
$^2$Harvard Quantum Initiative, Harvard University, Cambridge, MA 02138, USA}

\date{\today}

\begin{abstract}
Neutral atom arrays driven into Rydberg states constitute a promising 
approach for realizing programmable quantum systems. 
Enabled by strong interactions associated with Rydberg blockade, they allow for simulation of complex spin models and quantum dynamics. We introduce a new Floquet engineering technique for systems in the blockade regime that provides control over novel forms of interactions and entanglement dynamics in such systems. Our approach is based on time-dependent control of Rydberg laser detuning and leverages perturbations around periodic many-body trajectories as resources for operator spreading. These time-evolved operators are utilized as a basis for engineering interactions in the effective Hamiltonian describing the stroboscopic evolution. As an example, we show how our method can be used to engineer strong spin exchange, consistent with the blockade, in a one-dimensional chain, enabling the exploration of gapless Luttinger liquid phases. In addition, we demonstrate that combining gapless excitations with Rydberg blockade can lead to dynamic generation of large-scale multi-partite entanglement. Experimental feasibility and  possible generalizations are discussed.  
\end{abstract}

\maketitle

{\it Introduction.---} 
Programmable quantum simulators provide unique insights into complex many-body systems. They can be used for explorations of strongly correlated quantum 
phases of matter~\cite{mazurenko2017,ebadi2021,semeghini2021,semeghini2021,chen2023,feng2023}, non-equilibrium 
quantum dynamics~\cite{langen2015,schreiber2015,kaufman2016,zhang2017,bernien2017,zhang2017_DTC,bluvstein2021,bornet2023,hines2023,manovitz2024quantum}, many-body entanglement~\cite{fukuhara2015,islam2015,kaufman2016,omran2019,brydges2019,joshi2023}, and quantum metrology~\cite{schleier2010,pedrozo2020,colombo2022,bornet2023,eckner2023,hines2023}.
Neutral atom arrays are a promising approach to realizing programmable quantum simulators~\cite{bernien2017,browaeys2020,labuhn2016,morgado2021}, 
where tunable atom trapping geometry along with strong interactions
resulting in Rydberg blockade allow one to generate a variety of strongly correlated spin models. Coherent laser excitation into Rydberg states  
generates dynamics within the accessible Hilbert space, similar to that 
provided by a global transverse field in the Ising model. 
These constrained dynamics
result in new physical phenomena such as quantum many-body scars~\cite{bernien2017,turner2018,bluvstein2021}, which evade thermalization starting from certain product initial states.
At the same time, 
extending  the toolbox of Rydberg quantum simulation in the blockade regime to dynamical generators beyond simple transversal fields is an open challenge. This is 
important,  
for instance, for realizing  spin liquids~\cite{semeghini2021,Verresen2021_RSL,Verresen2022_Sls,tarabunga2022_RSL} and lattice gauge theories~\cite{surace2020_lgt,Samajdar2021_kagome,samajdar2023_z2,celi2020_lgt,Homeier_2023vt}, for steering the dynamics of many-body states via counterdiabatic terms~\cite{demirplak2003_adiabatic,Sels2017_CDD,ljubotina2022}, realizing new types of quantum optimization algorithms~\cite{pichler2018quantumoptimizationmaximumindependent,cain2023,nguyen2023gadgets} and for recent efforts to generate metrologically useful entanglement in systems with quantum many-body scars~\cite{desaules2022,dooley2023}.

Motivated by these considerations, in this Letter we introduce a technique for Floquet engineering~\cite{goldman2014_floquet,bukov2015universal,eckardt2017_floquet,aidelsburger2013_hofstadter,miyake2013_harper,Jotzu2014ug,flaeschner2016_floquet,meinert2016_floquet,geier2021_floquet,scholl2022_floquet} that employs time-dependent control to realize effective Rydberg-blockaded models with versatile interactions.
While Floquet engineering is widely utilized for interacting spin systems~\cite{wahuha1968,wei2018_nmr,choi2020_pulses,zhou2020_metro,zhou2023_hameng,geier2024time}, conventional techniques rely on local Pauli frame transformations that generally violate the blockade constraint.
Our approach, illustrated in \figc{fig:1}{b}, leverages driven, periodic many-body trajectories originally discovered in the context of stabilizing quantum many-body scars~\cite{bluvstein2021,maskara2021,hudomal2022_drive}.
The complex micromotion of these trajectories serves as a resource for programmable Hamiltonian engineering, since perturbations applied during the periodic drive act at stroboscopic times via an effective Hamiltonian generated by time-evolved operators~\cite{maskara2021}.
The resulting class of time-evolved operators forms a basis  for the  realization of novel interactions with tunable coefficients,
which are not accessible in the static native Hamiltonian.
Using this approach, we show how blockade-consistent spin exchange interactions can be engineered, enabling the investigation of gapless phases with emergent particle number conservation~\cite{verresen2019}. Moreover, in this new regime, we demonstrate the dynamic generation of structured multi-partite entanglement from Néel ($\neel$) product initial states and explain this effect in terms of the dynamics of domain walls.

{\it Hamiltonian engineering.---} The key idea of this work can be understood by  considering driven PXP model illustrated in \figc{fig:1}{a}, which describes a one-dimensional atom chain with periodic boundary conditions, driven into the Rydberg state
 with fixed Rabi  frequency $\Omega$ under idealized nearest-neighbor Rydberg blockade, with time-dependent global detuning $\Delta(t)$:
\begin{align}
    H(t) &=  \frac{\Omega}{2} \Hpxp - \Delta(t) N, \label{eqn:native} \\
    \Hpxp &= \sum_i P_{i-1} \sigma_i^x P_{i+1}, \; \; N = \sum_i n_i.
\end{align}
Here the operators $P_i =\frac{\mathds{1}+\sigma_i^z}{2} = |\circ\rangle_i\langle\circ|_i$ and $n_i =\frac{\mathds{1}-\sigma_i^z}{2}= |\bullet\rangle_i\langle\bullet|_i$ project site $i$ onto ground ($\circ$) and Rydberg ($\bullet$) states respectively, while $\sigma_i^x=|\circ\rangle_i\langle\bullet|_i + |\bullet\rangle_i\langle\circ|_i$ generates Rabi oscillations. 
\begin{figure}
    \centering
    \includegraphics[width=\linewidth]{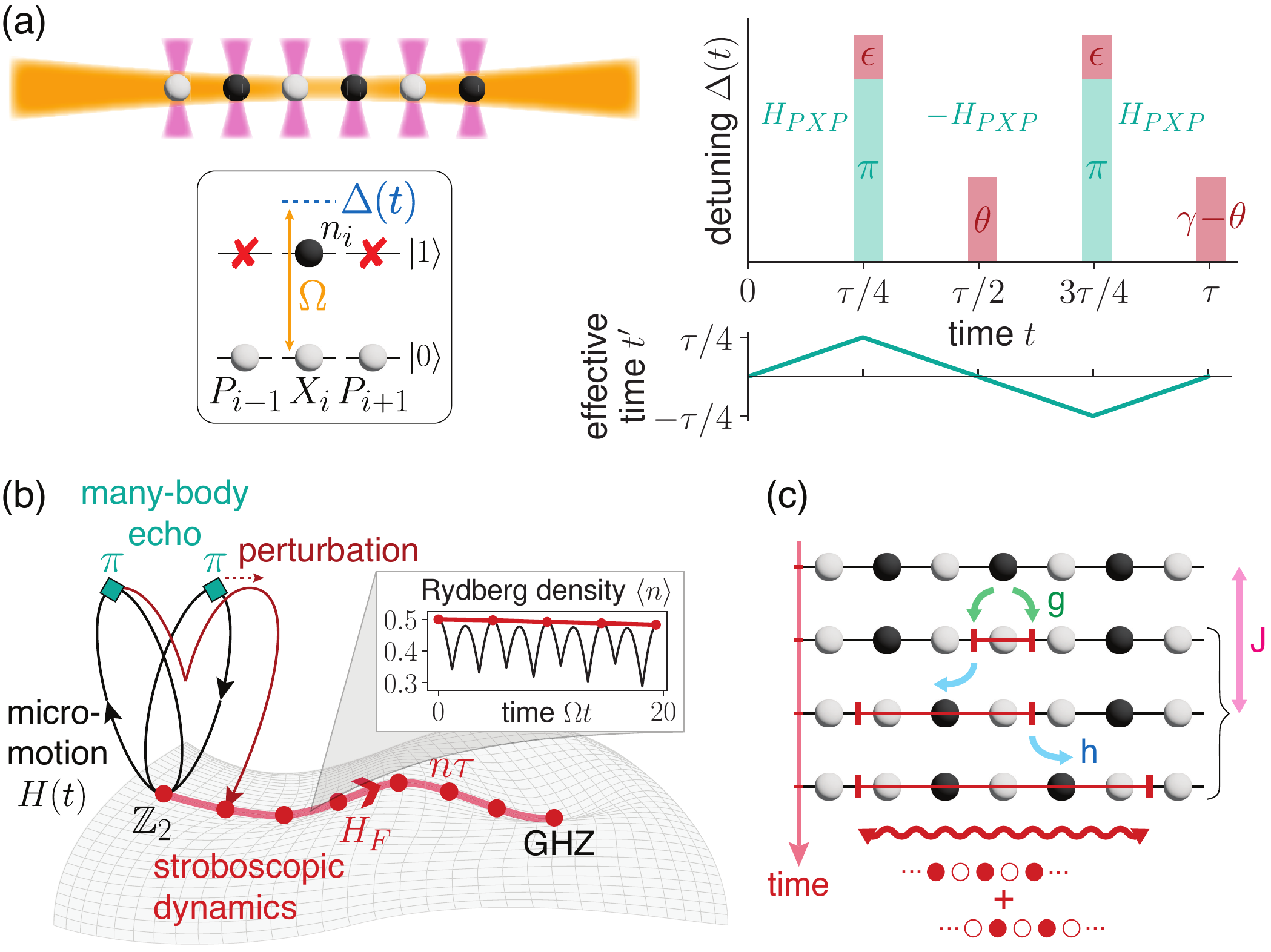}
    \caption{\textbf{Hamiltonian engineering.} \textbf{(a)} We consider the driven PXP model as an approximate description of Rydberg atoms in optical tweezers, with Rabi frequency $\Omega$ and $\tau$-periodic time-dependent global detuning $\Delta(t)$.
    Our protocol consists of $\pi$-pulses that realize a many-body echo (green) and deliberately placed perturbations (red). The echo realizes evolution under $\Hpxp$ for effective time $-\frac{\tau}{4} \leq t^{\prime} \leq \frac{\tau}{4}$, returning to $t^{\prime}=0$ at stroboscopic times.
    \textbf{(b)} 
    The resulting dynamics perturbs around the periodic trajectories of the many-body echo, illustrated as micromotion  in the full blockade-constrained Hilbert space. 
    At stroboscopic times $n\tau$, an effective Floquet Hamiltonian $H_F$ generates evolution within a constant energy submanifold.
    In our main application, the stroboscopic dynamics evolves a N\'eel ($\neel$) initial state towards a highly entangled GHZ state. Inset: Micromotion (black) and stroboscopic dynamics (red) of the Rydberg density $\braket{n}$ on a $L=16$ periodic chain. Despite large oscillations within Floquet cycles, the stroboscopic evolution approximately conserves $\braket{n}$.
    \textbf{(c)} Our approach realizes effective models $H_F$ with tunable control over the dynamics of domain wall excitations on top of the N\'eel order.
    This includes a chemical potential $J$, blockade-consistent spin exchange interactions $h$ that act as two-site hopping for domain walls, as well as creation/annihilation terms $g$ of domain wall pairs.
    The regime of small $g$ provides a mechanism for generating long-range multi-partite entanglement via growing superpositions of alternate $\neel$ orders. 
    }
    \label{fig:1}
\end{figure}
Our objective is to realize off-diagonal number conserving processes beyond single spin flips, utilizing the intrinsic controls $\Omega$ and $\Delta(t)$. In a conventional (static) approach ($\Delta(t)=\Delta$) to this problem, one  typically relies on large detuning $\Delta \gg \Omega$, where multi-body spin flips emerge perturbatively in powers of $(\Omega/\Delta)^n$. The relative strengths of such processes is generally weak and cannot be  tuned independently. 

Our dynamic protocol circumvents this restriction by modulating laser detuning $\Delta(t)$ and leveraging  a many-body spin echo. Specifically, 
since $\Hpxp$ anti-commutes with the operator $\prod_i \sigma_i^z=e^{i \pi N}$, 
$\pi$-pulses of the global detuning effectively reverse its sign~\cite{turner2018,maskara2021}: $e^{i\pi N} \Hpxp e^{i\pi N} = -\Hpxp$.
This property enables a dynamical decoupling in the strongly interacting PXP model by means of a simple pulse sequence ($n \in \mathbb{N}$),
\begin{equation} \label{eqn:many_body_echo}
\begin{split}
    \Delta_0(t) = \pi \; \sum_n \delta\bigl(t-\frac{\tau}{4}-n \frac{\tau}{2}\bigr).
\end{split}
\end{equation}
Within each Floquet period $\tau$, the system evolves forward under $\Hpxp$ for $\tau/4$, backward for $\tau/2$, and forward again for $\tau/4$.
At a given time $t \leq \tau$, the system has thus undergone an effective evolution by $\Hpxp$ for a time $t^{\prime}= t^{\prime}(t) =||t-\tau/4|-\tau/2|-\tau/4$, where $t^\prime \in [-\frac{\tau}{4},\frac{\tau}{4}]$, as shown in \figc{fig:1}{a} (see Supplemental Material~\cite{supplement}, \sect{sec:many-body-echo}). As $t^\prime (\tau)=0$, the system exhibits periodic revivals at stroboscopic times $n\tau$.

To generate a nontrivial effective evolution,  we utilize complex micromotion along the periodic trajectory, as illustrated in \figc{fig:1}{b}. Specifically, we introduce global detuning perturbations 
consisting of $\tau$-periodic discrete pulses around the echo protocol of \eq{eqn:many_body_echo}, $\Delta(t)=\Delta_0(t)+\tilde{\Delta}(t)$ with $\tilde{\Delta}(t)=\sum_{j,n} \tilde{\Delta}_j \, \delta\left(t-t_j-n\tau\right)$.
At stroboscopic times, these perturbations translate into evolution under a static, local effective Floquet Hamiltonian $H_F$, which holds up to an exponentially long prethermal timescale $T_p \gtrsim (\tau/|\tilde{\Delta}|) e^{c_p/|\tilde{\Delta}|}$, where $|\tilde{\Delta}|=\sum_j |\tilde{\Delta}_j|$ and $c_p > 0$~\cite{abanin2017,abanin2017_2}.
In the interaction picture with respect to the perfect echo evolution for $\tilde{\Delta}=0$, we obtain the leading contributions to $H_F$ through a Floquet-Magnus expansion~\cite{else2017,maskara2021} (see~\cite{supplement}, \sects{sec:perturbations}{sec:effective_ham}), resulting in $H_F = \sum_{n=0}^\infty H_F^{(n)}$, with
\begin{equation} \label{eqn:FM_expansion}
\begin{split}
    H_F^{(0)} = -\sum_j \frac{\tilde{\Delta}_j}{\tau} \tilde{N}(t^{\prime}_j), \;
    H_F^{(1)} = \sum_{j>\ell} \frac{\tilde{\Delta}_j \tilde{\Delta}_\ell}{2i\tau} \left[ \tilde{N}(t^{\prime}_j), \tilde{N}(t^{\prime}_\ell)\right].
\end{split}
\end{equation} 
We note that $H_F^{(n)}$ are constructed from Rydberg number operators conjugated by evolution under $\Hpxp$ for the effective times $t^{\prime}_j = t^\prime (t_j)$ of the echo protocol,
$\tilde{N}(t^\prime) \equiv e^{it^\prime\frac{\Omega}{2}\Hpxp} N e^{-it^\prime\frac{\Omega}{2}\Hpxp}$.
Consequently, operator spreading under the micromotion generated by $\Hpxp$ induces interactions in the form of $n$-nested commutators $(t^{\prime})^n
[\Hpxp,...,[\Hpxp,N]...]$ in $\tilde{N}(t^{\prime})$ and thus $H_F$. 

Coefficients of these terms in $H_F$ are controlled by the locations $t_j$ and weights $\tilde{\Delta}_j$ of the pulses. Pulses with $t^{\prime}_j=0$
couple to the bare Rydberg number operator $N$. 
Further, symmetric weights of pulses at $\pm t^\prime_j$
ensures that all terms containing an \textit{odd} number of commutators vanish in $H_F^{(0)}$, which thus conserves Rydberg number parity. 
Hence, for small $t^{\prime}_j$ the leading non-trivial contribution to $H_F^{(0)}$ appears at order $(t^{\prime}_j)^2$, which contains nearest-neighbor pair flips.
Single spin flips appear only in $H_F^{(1)}$. 
Based on this intuition, we introduce the following detuning perturbations, parameterized in terms of the (dimensionless) variables $\gamma,\theta,\epsilon$,
see \figc{fig:1}{a},
\begin{equation} \label{eqn:time_dep_detuning}
\begin{split}
    \tilde{\Delta}(t) = &(\gamma-\theta) \;\sum_n\delta\left(t-n\tau\right) + \theta\; \sum_n\delta\bigl(t-\frac{\tau}{2}-n\tau\bigr)\\
    + &\epsilon \; \sum_n\delta\bigl(t-\frac{\tau} {4}-n\tau\bigr) + \epsilon \; \sum_n\delta\bigl(t-\frac{3\tau}{4}-n\tau\bigr).
\end{split}
\end{equation}
Here, the first two pulses both occur at effective time $t^\prime=0$, while the latter occur at $t^\prime = \pm \frac{\tau}{4}$.
Inserting into \eq{eqn:FM_expansion} and expanding $\tilde{N}(\tau/4)$ perturbatively for short periods $\frac{\Omega\tau}{4} \ll 1$, we obtain an approximate closed-form expression for the effective Hamiltonian,
\begin{equation} \label{eqn:effective_model}
\begin{split}
    &H_F \approx -J\; N - h\; \Hpxyp + g\; \Hpxp + \frac{h}{4}\; H_{ZIZ}, \\
    &J = \frac{(\gamma+2\epsilon)}{\tau} - \frac{3\epsilon \Omega^2 \tau}{32},\; h = - \frac{\epsilon \Omega^2 \tau}{32},\; g = - \frac{\epsilon(\theta+\epsilon)\Omega}{8}.
\end{split}
\end{equation}
Here, we have kept all terms to quadratic order in $\Omega\tau,\epsilon,\gamma,\theta$ (see details in~\cite{supplement}, \sect{sec:effective_ham});
$\Hpxyp \equiv \frac{1}{2} \sum_i P_{i-1} \left( \sigma_i^x \sigma_{i+1}^x + \sigma_i^y \sigma_{i+1}^y \right) P_{i+2}$ is a blockaded nearest-neighbor spin-exchange interaction and $H_{ZIZ} \equiv \sum_i \sigma_i^z \sigma_{i+2}^z$.  
\begin{figure}
    \centering
    \includegraphics[width=\linewidth]{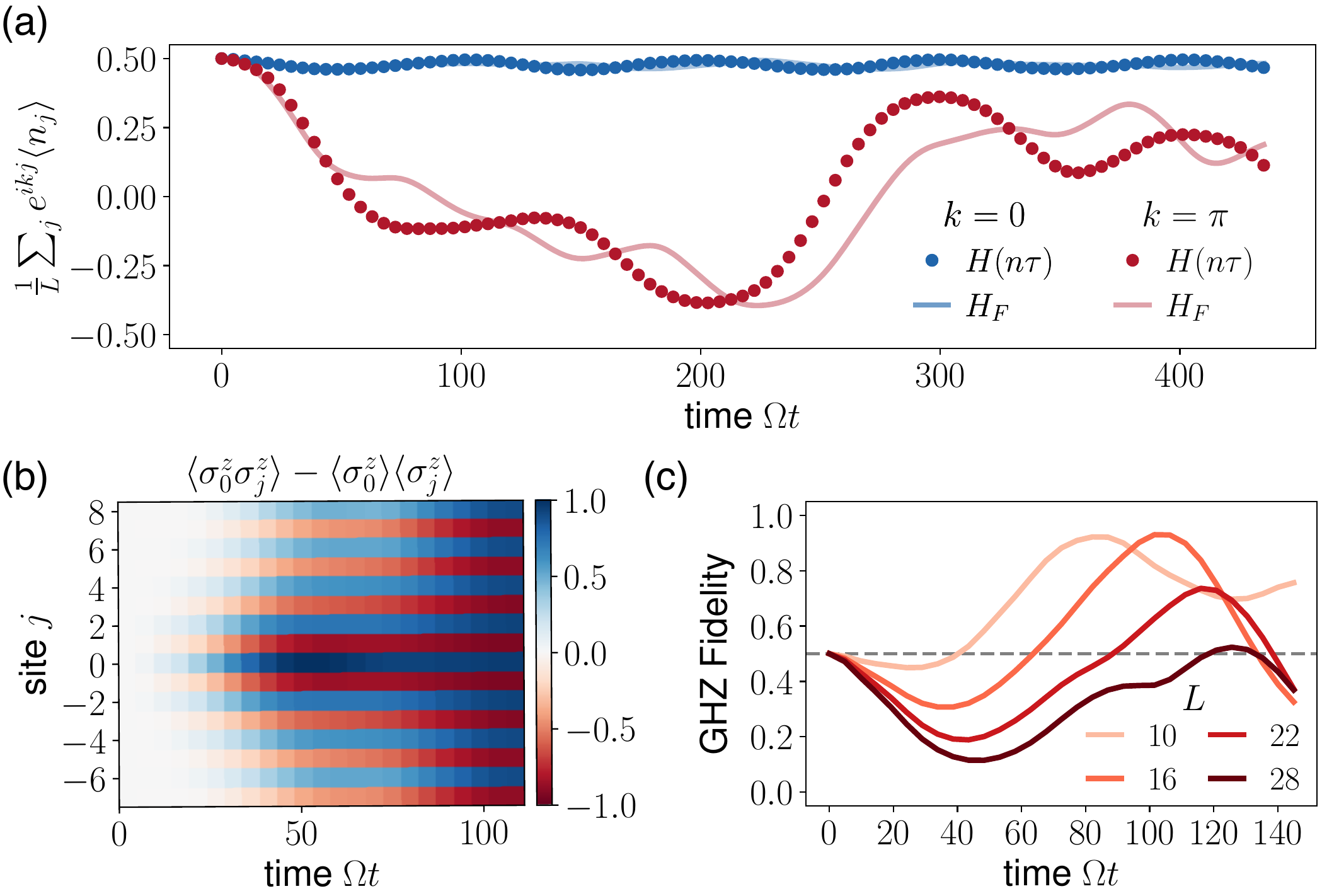}
    \caption{\textbf{Entanglement dynamics.} \textbf{(a)} Stroboscopic evolution of a $L=16$ periodic chain under a drive with period $\tau=2\pi/1.3$ and perturbations set to $\epsilon=-0.45$, $\gamma=1.0, \theta=0.15$. Starting from a N\'eel initial state, the Rydberg number density (blue) remains high, while the $\neel$ order (red) washes out, in qualitative agreement with evolution under the effective Hamiltonian \eq{eqn:effective_model}. 
    \textbf{(b)} Dynamics under this drive generates spreading, connected $\sigma^z$-correlations with antiferromagnetic (AFM) spatial profile. \textbf{(c)} The dynamics realizes an AFM GHZ state for $L=16$ once the connected $\sigma^z$-correlations peak near unity at all distances. As system size increases, the peak in GHZ fidelity shifts approximately linearly with $L$, with decreasing height.
    }
    \label{fig:2}
\end{figure}
Our Floquet protocol provides flexible relative tunability of the coefficients $J,h,g$ by controlling the period $\Omega\tau$ and parameters $\epsilon, \gamma, \theta$. Thus, our construction extends the capabilities of the Rydberg quantum simulator to blockade models with independent control over quasiparticle number conservation, motion, and creation/annihilation processes. In particular, this enables access to exchange-dominated regimes $h \gtrsim g$, which we explore in the following.

{\it Controlled multi-partite entanglement.---} 
We consider dynamics starting from a Néel state, $\ket{\Psi(t=0)}=\ket{\mathbb{Z}_2}$. Excitations on top of this state can be viewed as domain walls, created in pairs by $\Hpxp$, see \figc{fig:1}{c};  $J$ sets a chemical potential and $\Hpxyp$ generates (two-site) hopping of domain walls. We fix the parameters of the Floquet drive to $\Omega\tau/2\pi = 0.77$ (close to the scar period of the bare PXP model~\cite{bernien2017}), and $\epsilon=-0.45, \gamma=1.0, \theta=0.15$. This translates to an effective model of \eq{eqn:effective_model} with $J\approx0.225, h\approx0.068, g\approx-0.017$, where the rate of domain wall hopping is stronger than of creation/annihilation. As shown in \figc{fig:1}{b}, the average Rydberg density $\braket{n(t)}$ varies rapidly during micromotion, but evolves only slowly at stroboscopic times. Even though the parameters perturbing the echo are sizeable, the effective Hamiltonian \eq{eqn:effective_model} nonetheless provides a good description of the stroboscopic evolution as demonstrated in \figc{fig:2}{a}.
Interestingly, while the density of Rydberg excitations remains high, the staggered magnetization $\sum_j (-1)^j\braket{\sigma^z_j(t)}$ washes out. As this happens, the system develops growing connected correlations $\braket{\sigma^z_i(t)\sigma^z_j(t)}-\braket{\sigma^z_i(t)}\braket{\sigma^z_j(t)}$; \figc{fig:2}{b}. Within the correlated region, a superposition of alternate $\mathbb{Z}_2$ orders emerges, akin to patches of GHZ states.
Once these large scale fluctuations reach the system size (here: $L=16$), the state develops large overlap with an antiferromagnetic GHZ state, defined as $|\text{GHZ} \rangle = \frac{1}{\sqrt{2}} \left( |\neel\rangle + e^{i \phi} |\neelp \rangle \right)$, which we quantify via the fidelity $\max\limits_{\phi}{|\langle \text{GHZ}|\Psi(t)\rangle|^2}$
~\cite{monz2011,omran2019}, see \figc{fig:2}{c}.
As system size increases, the time of maximum GHZ fidelity changes roughly linearly in $L$, suggesting that the relevant processes are not exponentially suppressed. 
At the same time, the corresponding peak height decreases with $L$, indicating that correlations are less likely to reach the full system size.

These results can be understood based on the effective model \eq{eqn:effective_model}, as sketched in \figc{fig:1}{c}: $g\Hpxp$ slowly creates a superposition of the initial $\mathbb{Z}_2$ state with states containing pairs of domain walls, which then move rapidly via the strong $h\Hpxyp$ interaction. The propagating pair carries a growing string of the alternate $\neelp$ order, thus producing antiferromagnetic GHZ-like correlations. On a periodic chain, the pair may re-annihilate at the antipodal point to form the state $\ket{\text{GHZ}}$.
Based on this picture, we expect that the coherent spread of correlations persists over a timescale $t^*\sim 1/g$, beyond which it is interrupted by the emergence of additional domain walls. The size of the GHZ-like patch $l^*$ is determined by the rate of hopping $v^* \sim h$ within this timescale, $l^*=v^*t^*\sim h/g$.
For $g\sim 1/L$, this size is $l^* \sim L$, and a full chain GHZ state may form in a time linear in system size.

We confirm these predictions using numerical analysis of the Quantum Fisher Information (QFI) density, which quantifies the metrological potential of the pure state $\ket{\Psi(t)}$, evolving under $H_F$, with respect to a staggered $z$-field,
$F_Q/L \equiv \frac{1}{L} \text{Var}\left( \textstyle\sum_j (-1)^j \sigma_j^z \right)^2_{|\Psi(t)\rangle}$.
In particular, a QFI density $F_Q/L > m$ implies at least $(m+1)$-body entanglement~\cite{hyllus2012}, such that $F_Q/L>1$ indicates non-classical correlations and $F_Q/L=L$ corresponds to a maximally entangled GHZ state, which saturates the Heisenberg limit. In \figc{fig:3}{b}, iTEBD simulations of an infinite chain demonstrate that the QFI obeys the predicted scalings for the formation time $t^*$ and maximum size $l^*$ of the multi-partite entangled regions.

\begin{figure}
    \centering
    \includegraphics[width=\linewidth]{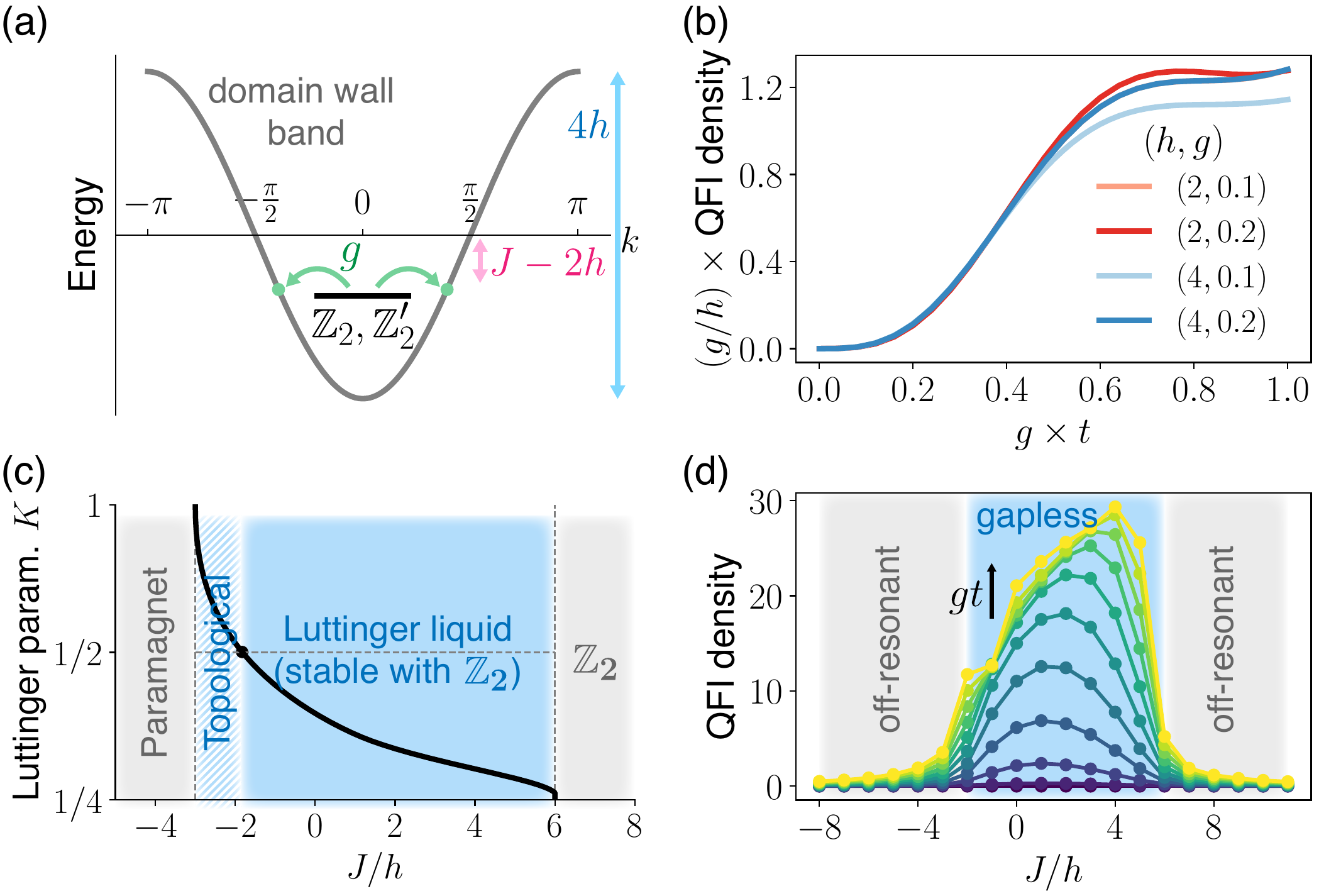}
    \caption{\textbf{Blockaded spin-exchange and approximate $U(1)$ symmetry.} \textbf{(a)} The effective Hamiltonian of \eq{eqn:effective_model} for 
    $g/h \lesssim 1$ weakly couples the N\'eel states 
    to pairs of domain walls  with momentum $\pm k$ and quasiparticle dispersion $\varepsilon_k= -2h\cos{k}$. Depending on the energy offset between Néel states and the dispersive band, this coupling can be on- or off-resonant.
    \textbf{(b)} Dynamics under $H_F$ for small $g$ generates a maximum QFI density (see text) of order $\sim h/g$ within a time $t^*\sim 1/g$. This is seen through a scaling collapse for the dynamics of an infinite size chain simulated via iTEBD for different values of $h,g$ at fixed $J=2h$.
    \textbf{(c)} Phase diagram of $H_F$ at the integrable point $g=0$, with gapped paramagnetic and $\neel$ phases, as well as 
    a Luttinger liquid that is stable for $K<1/2$ (in the presence of $\mathbb{Z}_2$ number parity), and unstable for $K>1/2$ (towards a gapped topological state~\cite{verresen2019,fendley2004}).
    The Luttinger parameter $K$ (black line) is obtained from the Bethe ansatz solution. 
    \textbf{(d)} iTEBD simulation of the QFI density from a $\mathbb{Z}_2$ initial state for weakly broken integrability at small $g$. The QFI functions as a dynamical probe of the transition between $\mathbb{Z}_2$ phase and Luttinger liquid in c), and grows at a rate tied to the domain wall dispersion. 
    }
    \label{fig:3}
\end{figure}

{\it Luttinger liquid dynamics---} 
Due to the small value of $g$, the entanglement features observed in the previous section may be understood as a dynamical probe of the low energy properties of the constrained model \eq{eqn:effective_model} with $U(1)$ symmetry at $g=0$. In particular, $H_F|_{g=0}$ is known to be integrable~\cite{alcaraz1999}, and single domain walls form a band of quasiparticles with dispersion $\varepsilon_k = -2h \cos(k)$. The Néel $\mathbb{Z}_2$-states are offset from the center of this band by an energy $J-2h$ due to chemical potential and $H_{ZIZ}$ term, see \figc{fig:3}{a}.
The ground state phase diagram of this model, which we calculate from Bethe ansatz integral equations in~\cite{supplement}, \sect{sec:effective_model_phases}, is shown in \figc{fig:3}{c}. For $J/h \in (-3,6)$, the system is in a Luttinger liquid phase with gapless domain wall excitations. 
Outside this regime, the ground state transitions into a gapped paramagnetic ($J/h<-3$) or $\neel$ ($J/h>6$) phase.

From this phase diagram, we see that the protocol of the previous section corresponds to a quantum quench of the $\mathbb{Z}_2$ state into the Luttinger liquid phase. For small $g$, the initial $\mathbb{Z}_2$ state couples \textit{resonantly} to pairs of domain walls with momenta $\ket{k,-k}$ such that $J-2h+2\varepsilon_k = \mathcal{O}(g)$, which mediate the growing entanglement as described above. 
In particular, this coupling is proportional to the group velocity of the cosine dispersion, $\braket{k,-k|gH_{PXP}|\mathbb{Z}_2} \sim 2 g \sin(k)$ (see derivation in~\cite{supplement}, \sect{sec:domain_wall_dynamics}).
For $J/h$ outside the Luttinger liquid phase, domain walls can only be created virtually (i.e. off-resonantly), suppressing the growth of the QFI. As a consequence, dynamics of multipartite entanglement~\cite{weidinger2018}, quantified by the QFI density, directly probes the transition from the gapped $\mathbb{Z}_2$ symmetry-breaking phase to the gapless Luttinger liquid, with its early time growth reflecting the quasiparticle group velocity. 
We confirm this prediction numerically by computing the dynamics of the QFI under $H_F$ of \eq{eqn:effective_model} using iTEBD for different values of $J/h$, see \figc{fig:3}{d}. The growth of the QFI accurately captures the phase boundary at $J/h=6$ and shows an enhanced rate towards the center of the Luttinger liquid phase. 
We have also verified these features 
in a direct simulation of the Floquet protocol for finite systems, see~\cite{supplement}, \sect{sec:dynamical_probe_gapless}.

{\it Robustness and Implementation in Rydberg arrays.---} 

Successfully realizing and controlling $H_F$ in \eq{eqn:effective_model} relies on the expansions in small perturbations $\epsilon,\gamma,\theta$ and drive periods $\Omega\tau$, which also tune the rate of dynamics. Thus, an optimal choice of drive parameters must consider the trade-off between a high-accuracy target Hamiltonian and the realistic constraint of observing significant dynamics within finite coherence time of an experimental device, an interplay we explore in~\cite{supplement}, \sect{sec:performance_floquet}. 
In addition, in experiment, Rydberg atoms interact via van der Waals interactions $V_{\mathrm{vdW}}=\Omega \sum_{ij}\bigl(\frac{R_b}{|i-j|}\bigr)^6n_i n_j$, with blockade radius $R_b$.
To demonstrate that our approach applies qualitatively, we construct a  pure spin exchange model ($-2\epsilon=\gamma=2\theta \Rightarrow J=g=0, h\neq 0$) and 
numerically evaluate the quantum walk of a single Rydberg excitation resulting from Floquet evolution including $V_{\mathrm{vdW}}$ at $R_b=1.5$.
We further add a small constant detuning to mitigate the long range tail of $V_{\mathrm{vdW}}$~\cite{bluvstein2021} 
and use Gaussian pulse profiles with finite width, 
suitable for
realistic control hardware with 
limited rate of detuning modulation;
\figc{fig:4}{a}.
Comparison with the corresponding PXP result in \figc{fig:4}{b} shows very good qualitative agreement over an extended duration. Quantitatively, the effective hopping is even stronger in the experimentally relevant scenario, likely due to residual contributions from finite pulse width and long range tails. A detailed analysis of these effects is left for future work.

\begin{figure}
    \centering
    \includegraphics[width=\linewidth]{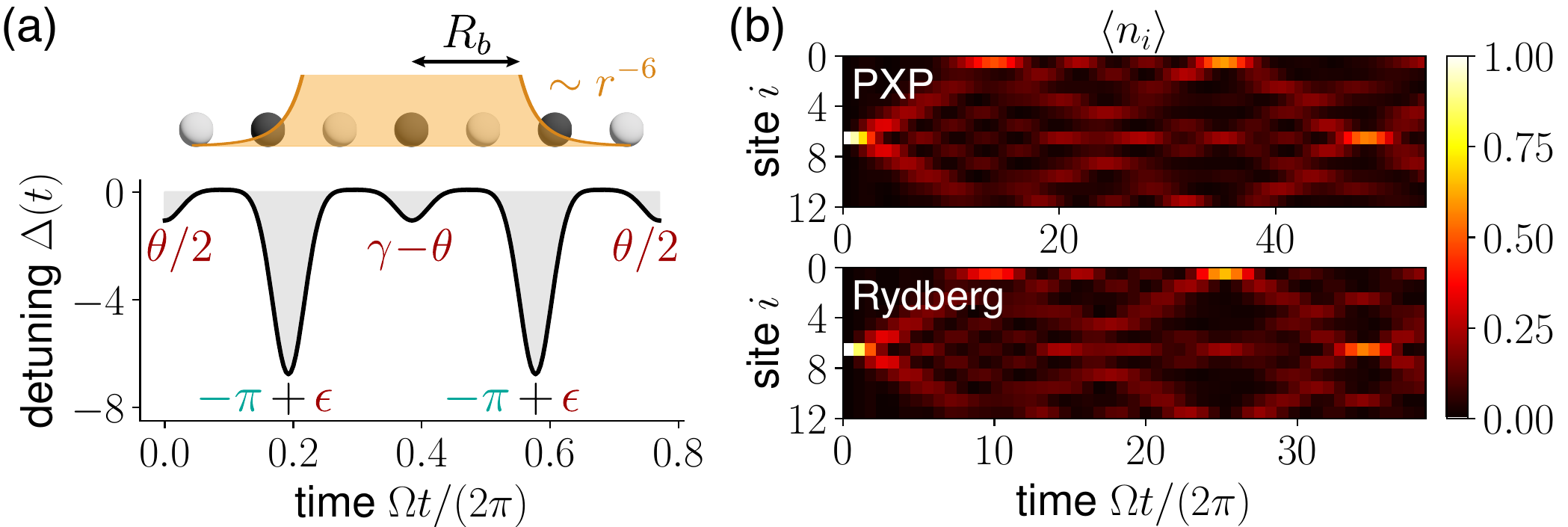}
    \caption{\textbf{Experimental applicability.} \textbf{(a)} In experiment, Rydberg atoms at distance $r=|i-j|$ interact via $V_{\mathrm{vdW}}$ with strength $\Omega (R_b/r)^6$; here, blockade radius $R_b=1.5$. In implementing the drive in (b), we use realistic Gaussian pulse profiles with width $w=0.046\tau$, ensuring each pulse integrates to the desired detuning (we apply $-\pi$-pulses to avoid blockade violations). A small constant detuning $\delta_{\text{MF}}/\Omega = 0.09$ is added to mitigate the long range tail of $V_{\mathrm{vdW}}$~\cite{bluvstein2021}.
    \textbf{(b)} We benchmark Floquet evolution of the full Rydberg Hamiltonian under finite-width pulses shown in (a), against the idealized PXP model driven by infinitely sharp pulses. We apply a drive $\epsilon=-\gamma/2=-\theta=0.45$ that generates an effective Hamiltonian with $h=-\frac{\epsilon\Omega^2\tau}{32}$, $J=g=0$ to engineer a single-particle quantum walk on an $L=12$ periodic chain. The dynamics of local Rydberg occupations $\braket{n_i}$ are consistent with this target; the effective hopping is slightly stronger in the Rydberg model.}
    \label{fig:4}
\end{figure}

{\it Discussion \& Outlook.---}
We have introduced a Floquet protocol for systems of Rydberg atoms that exploits periodic trajectories of quantum states, enabling versatile Hamiltonian engineering. As an application, we realized models with emergent particle number conservation and dominant blockade-consistent exchange interactions in one dimension, 
exploring previously inaccessible gapless Luttinger liquid phases. In particular, we found that combining Rydberg blockade and gapless domain wall excitations leads to the generation of long-range, multi-partite entanglement upon evolving $\mathbb{Z}_2$ product states. 

Although our discussion focuses on one-dimensional systems, generalization to other geometries is natural. For instance, models akin to \eq{eqn:effective_model} 
are relevant to Rydberg spin liquids in two dimensions~\cite{Verresen2022_Sls,tarabunga2022_RSL}, as well as to achieving quantum speedup in combinatorial optimization by enabling delocalization in the adiabatic algorithm~\cite{cain2023}.
Moreover, going beyond short evolution times of the operator $\tilde{N}(t)$ provides access to even higher-body spin interactions, an approach we employ in Ref.~\cite{feldmeier2024} to study the dynamics of two-dimensional lattice gauge theories.
We further emphasize that our Floquet scheme uses only simple global controls, but may be extended by incorporating site-resolved detuning fields~\cite{manovitz2024quantum}, which could allow exploration of chiral interactions~\cite{naveen2023}.
It would also be interesting to study the connection of our protocol with time-dependent methods for state preparation, such as counter-diabatic driving~\cite{demirplak2003,cepaite2023,schindler2024} and trajectory optimization~\cite{ljubotina2022}.
Finally, we note that our Floquet protocol can be extended to other quantum simulation platforms, contingent on time-dependent control to engineer non-trivial periodic trajectories, as is available in dipolar interacting systems~\cite{wahuha1968,wei2018_nmr,choi2020_pulses,zhou2020_metro,
geier2024time}, trapped ions~\cite{gaerttner2017_tr,gilmore2021_tr}, neutral atoms~\cite{linnemann2016_tr,Colombo2022_tr}, or superconducting devices~\cite{blok2021_scramble,mi2022_tr,Braumuller2022_tr}.

\emph{Acknowledgments.---}We thank Gefen Baranes, Dolev Bluvstein, Pablo Bonilla, Madelyn Cain, Simon Evered, Alexandra Geim, Andi Gu, Marcin Kalinowski, Nathaniel Leitao, Sophie Li, Tom Manovitz, Varun Menon, Simone Notarnicola, Hannes Pichler, Maksym Serbyn, and Peter Zoller for stimulating discussions.
Matrix product state simulations were performed using the TeNPy package~\cite{hauschild:2018}.
We acknowledge financial support from the US Department of Energy (DOE Gauge-Gravity, grant number DE-SC0021013, and DOE Quantum Systems Accelerator, grant number DE-AC02-05CH11231), the National Science Foundation (grant number PHY-2012023), the Center for Ultracold Atoms (an NSF Physics Frontiers Center), the DARPA ONISQ program (grant number W911NF2010021), the DARPA IMPAQT program (grant number HR0011-23-3-0030), and the Army Research Office MURI (grant number W911NF2010082).
N.U.K. acknowledges support from The AWS Generation Q Fund at the Harvard Quantum Initiative.
N.M. acknowledges support by the Department of Energy Computational Science Graduate Fellowship under award number DE-SC0021110.
J.F. acknowledges support from the Harvard Quantum Initiative Postdoctoral Fellowship in Science and Engineering.

\bibliography{refs}

\clearpage

\setcounter{equation}{0}
\setcounter{page}{1}
\setcounter{figure}{0}
\renewcommand{\thepage}{S\arabic{page}}  
\renewcommand{\thefigure}{S\arabic{figure}}
\renewcommand{\theequation}{S\arabic{equation}}
\onecolumngrid
\begin{center}
\textbf{Supplemental Material:}\\
\textbf{Floquet engineering of interactions and entanglement \\ in periodically driven Rydberg chains}\\ \vspace{10pt}
Nazl\i \ U\u{g}ur K\"oyl\"uo\u{g}lu$^{1,2}$, Nishad Maskara$^{1}$, Johannes Feldmeier$^{1}$, and Mikhail D. Lukin$^{1}$ \\ \vspace{6pt}

\textit{\small{$^1$Department of Physics, Harvard University, Cambridge, MA 02138, USA \\ $^2$Harvard Quantum Initiative, Harvard University, Cambridge, MA 02138, USA}} 
\\

\vspace{10pt}
\end{center}
\maketitle
\twocolumngrid

\section{Floquet Engineering Protocol}
\label{sec:pulse_sequence_design}
In this section we discuss the Floquet engineering protocol in detail. We present the derivation and properties of the effective Floquet Hamiltonian for the stroboscopic dynamics of our time-dependent driving scheme.

\subsection{Many-body echo}
\label{sec:many-body-echo}
In the main text, we considered periodic many-body trajectories realized through a many-body echo, which involves evolving forward and backward under $\Hpxp$ for equal durations. 

Our ability to implement time-reversal of $\Hpxp$ relies on particle-hole symmetry: since $\Hpxp$ anti-commutes with the operator $\prod_i \sigma_i^z = e^{-i \pi N}$, its sign can be reversed using $\pi$-pulses of the global detuning, $e^{i \pi N}\Hpxp e^{i \pi N} = -\Hpxp$~\cite{turner2018,maskara2021}.
This feature enables dynamical decoupling of the strongly interacting, non-integrable PXP model using simple $\pi$-pulses, usually characteristic to dynamical decoupling of local, non-interacting fields. 

Utilizing this property, we consider time evolution $U_0(t) = \mathcal{T} e^{-i \int_0^t dt' H_0(t')}$ under the following drive:
\begin{align}
    H_0(t) &= \frac{\Omega}{2} \Hpxp - \Delta_0(t) N, \\
    \Delta_0(t) &= \pi \; \sum_n \delta\left(t -\frac{\tau}{4}-n\frac{\tau}{2}\right),
    \label{eqn:echo_drive}
\end{align}
which generates a Floquet unitary with period $\tau/2$:
\begin{equation}
\begin{split}
    \mathcal{X}_{\tau/2} \equiv U_0(\tau/2) &=e^{-i\frac{\tau}{4}\frac{\Omega}{2} \Hpxp} e^{i\pi N} e^{-i\frac{\tau}{4}\frac{\Omega}{2} \Hpxp} \\
    &= e^{i\pi N} e^{+i\frac{\tau}{4}\frac{\Omega}{2} \Hpxp} e^{-i\frac{\tau}{4}\frac{\Omega}{2} \Hpxp} \\
    &= e^{i\pi N}.
\end{split}
\end{equation}
This drive realizes a many-body echo at stroboscopic times $n \tau$ ($n \in \mathbb{Z}$), i.e. produces no effective dynamics:
\begin{equation}
\begin{split}
    U_0(\tau) = U_0(\tau/2)^2 = \mathcal{X}_{\tau/2}^2 = \mathbb{1}.
\end{split}
\end{equation}

However, during micromotion, i.e. in between stroboscopic times, the system follows a non-trivial periodic trajectory $U_0(t +n\tau)=U_0(t)$:
\begin{align}
    U_0(t) = 
    \begin{cases}
        e^{-it \frac{\Omega}{2} \Hpxp} & \text{if } 0 \leq t < \frac{\tau}{4}\\
        e^{i \pi N} e^{-i \left(\frac{\tau}{2}-t\right)\frac{\Omega}{2} \Hpxp} & \text{if } \frac{\tau}{4} \leq t < \frac{3\tau}{4} \\
        e^{-i \left(t-\tau\right) \frac{\Omega}{2} \Hpxp}  & \text{if } \frac{3\tau}{4} \leq t < \tau ,
    \end{cases}
    \label{eqn:frame_transformation}
\end{align}
which for $0 \leq t < \tau$ amounts to effective evolution by $\Hpxp$ for a time 
\begin{equation}
    t^{\prime} = t^{\prime}(t)=||t-\tau/4|-\tau/2|-\tau/4
    \label{eqn:effective_time_evol}
\end{equation}
with $t^{\prime} \in [-\frac{\tau}{4},\frac{\tau}{4}]$, followed by a detuning $\pi$-pulse if $\frac{\tau}{4}\leq t < \frac{3\tau}{4}$, and can be summarized as:
\begin{equation}
    U_0(t) = e^{-i \pi N \cdot \mathbb{1}_{\tau/4 \leq t <3\tau/4}} e^{-it^{\prime}\frac{\Omega}{2} \Hpxp}.
    \label{eqn:effective_frame_transform}
\end{equation}

In general, time evolution under an arbitrary drive can be analyzed
in the interaction picture with respect to the echo evolution $H_0(t)$, such that dynamics in the rotated ($\tilde{U}$) and 
laboratory ($U$) frames are related through the unitary transformation in \eq{eqn:frame_transformation}, $U(t) = U_0(t)^{\dagger}\tilde{U}(t) U_0(t)$.
Crucially, due to the many-body echo, the rotating and laboratory frames coincide at the end of each Floquet period $\tau$:
\begin{equation}
    U_F \equiv U(\tau) = U_0(\tau)^{\dagger} \tilde{U}(\tau) U_0(\tau) = \tilde{U}(\tau).
    \label{eqn:frame_coincide}
\end{equation}
Relying on this property, we now introduce perturbations around the many-body trajectory and analyze the effective Floquet dynamics in the rotating frame. 

\subsection{Controlled perturbations}
\label{sec:perturbations}

We consider perturbations around the many-body echo point $H_0(t)$, in the form of additional detuning pulses coupling to the global number operator $N$:
\begin{align}
    H(t) &= H_0(t) - \tilde{\Delta}(t) N \\
    \tilde{\Delta}(t) &=\sum_{j,n} \tilde{\Delta}_j \delta\left(t-t_j-n\tau\right)
    \label{eqn:detuning_pulses}.
\end{align}
The resulting time dynamics can be analyzed in a frame co-rotating with $H_0(t)$, as introduced in \sect{sec:many-body-echo}: $\tilde{H}(t) = - \tilde{\Delta}(t) U_0(t)^{\dagger} N U_0(t)$.
Following \eqs{eqn:effective_time_evol}{eqn:effective_frame_transform}, we introduce $N$ conjugated by evolution under $\Hpxp$ for effective time $t^{\prime}$,
\begin{equation}
    \begin{split}
    \tilde{N}(t^{\prime}) &\equiv e^{it^{\prime}\frac{\Omega}{2}\Hpxp}Ne^{-it^{\prime}\frac{\Omega}{2}\Hpxp} \\
    &= \sum_{n=0}^{\infty} \frac{(\Omega/2)^n}{n!} \left( t^{\prime}\right)^n \underbrace{[\Hpxp,...,[\Hpxp,}_{n \text{times}}N]...],
    \end{split}
    \label{eqn:rotated_N}
\end{equation}
to describe the number operator in the rotated frame: $U_0(t)^{\dagger} N U_0(t)=\tilde{N}(t^{\prime})$. 

Thus, we obtain the following dynamics in the rotated frame, in terms of pulses of $N(t^{\prime}_j)$ at locations $t_j$ with weights $\tilde{\Delta}_j$:
\begin{align}
    \tilde{H}(t) &= - \sum_{j,n} \tilde{\Delta}_j \delta\left(t-t_j-n\tau\right) \tilde{N}(t^{\prime}_j) \label{eqn:rotated_Ham_timedep}
    \\
    \tilde{U}(t) &= \mathcal{T} e^{-i \int_{0}^t \tilde{H}(t') dt'}.\label{eqn:rotated_unitary_timedep}
\end{align}
As per \eq{eqn:frame_coincide}, stroboscopic dynamics in the laboratory frame coincides with that in the rotated frame, and is given by the following Floquet unitary:
\begin{equation}
    U_F = \tilde{U}(\tau) = \prod_j e^{+i \tilde{\Delta}_j \tilde{N}(t_j)}.
\end{equation}

\subsection{Effective Hamiltonian}
\label{sec:effective_ham}
For small perturbations, the stroboscopic dynamics is well-described by a static effective Floquet Hamiltonian, $U_F \approx e^{-iH_F \tau}$, up to a prethermal timescale $T_p$ exponentially long in inverse perturbations strength, $T_p \gtrsim (\tau/|\tilde{\Delta}|) e^{c_p/|\tilde{\Delta}|}$, where $|\tilde{\Delta}|=\sum_j |\tilde{\Delta}_j|$ and $c_p > 0$~\cite{abanin2017,abanin2017_2}. We compute the leading order contributions to this effective Hamiltonian through a Floquet-Magnus expansion~\cite{else2017,maskara2021} of \eq{eqn:rotated_Ham_timedep}, resulting in $H_F = \sum_{n=0}^\infty H_F^{(n)}$, with
\begin{equation} \label{eqn:FM_expansion_app}
\begin{split}
    H_F^{(0)} &= \sum_j - \frac{\tilde{\Delta}_j}{\tau} \tilde{N}(t^{\prime}_j) \\
    H_F^{(1)} &= \sum_{j>\ell} \frac{\tilde{\Delta}_j \tilde{\Delta}_\ell}{2i\tau} \left[ \tilde{N}(t^{\prime}_j), \tilde{N}(t^{\prime}_\ell)\right].
\end{split}
\end{equation} 

In the specific parameterization of detuning perturbations provided in the main text,
\begin{align}
    \begin{split}
    \tilde{\Delta}(t) = &(\gamma-\theta) \;\sum_n\delta\left(t-n\tau\right) + \theta\; \sum_n\delta\bigl(t-\frac{\tau}{2}-n\tau\bigr)\\
    + &\epsilon \; \sum_n\delta\bigl(t-\frac{\tau} {4}-n\tau\bigr) + \epsilon \; \sum_n\delta\bigl(t-\frac{3\tau}{4}-n\tau\bigr),
\end{split}
\end{align}
pulses at $t_j=\left(0,\frac{\tau}{2}\right)$ couple to the bare Rydberg number operator $\tilde{N}(0)=N$, while pulses at $t_j=\left(\frac{\tau}{4},\frac{3\tau}{4}\right)$ couple to $\tilde{N}(\pm\frac{\tau}{4})$, generating the  
following Floquet unitary:
\begin{equation}
    U_F = e^{+i(\gamma-\theta)\tilde{N}(0)} e^{+i \epsilon \tilde{N}\left(-\frac{\tau}{4}\right)}  e^{+i\theta\tilde{N}(0)} e^{+i \epsilon \tilde{N}\left(\frac{\tau}{4}\right)}.
\end{equation}
The associated effective Floquet Hamiltonian is derived using \eq{eqn:FM_expansion_app}: 
\begin{align}
    \tau H_F^{(0)} &= - \gamma \tilde{N}(0) - \epsilon \left( \tilde{N}\left(\frac{\tau}{4}\right) + \tilde{N}\left(-\frac{\tau}{4}\right) \right) \label{eqn:FM_expansion_0} \\
\begin{split}
    2i\tau H_F^{(1)} &=  \epsilon (\gamma-2\theta)  \left[ \tilde{N}(0), \tilde{N}\left(-\frac{\tau}{4}\right) \right] +  \\
    & + \epsilon\gamma  \left[ \tilde{N}(0), \tilde{N}\left(\frac{\tau}{4}\right) \right] + \epsilon^2 \left[\tilde{N}\left(-\frac{\tau}{4}\right),\tilde{N}\left(\frac{\tau}{4}\right) \right].
\label{eqn:FM_expansion_1}
\end{split}
\end{align}

Symmetric weights of $\tilde{N}\left(\pm \frac{\tau}{4} \right)$ in $H_F^{(0)}$ ensures that all terms containing an $\emph{odd}$ number of commutators in \eq{eqn:rotated_N} vanish, which preserves Rydberg number parity. Moreover, for $\theta = \gamma/2$, this symmetry persists at all orders of the Floquet-Magnus expansion, in a weakly rotated basis. Specifically at this point, $\tilde{\Delta}(t)$ becomes periodic in $\tau/2$, and $\tilde{H}(t)$ possesses period-$\tau/2$ ``twisted time-translation symmetry'' with respect to the operator $\mathcal{X}_{\tau/2}$: $\tilde{H}(t+\tau/2)=\mathcal{X}_{\tau/2}^\dagger \tilde{H}(t) \mathcal{X}_{\tau/2}$. Using the formalism of Ref.~\cite{else2020}, the resulting Floquet unitary can be approximated as $U(\tau/2) \approx \mathcal{V} \mathcal{X}_{\tau/2} e^{-iD\tau/2} \mathcal{V}^\dagger$, and $U_F = {U(\tau/2)}^2 \approx \mathcal{V} e^{-iD\tau}\mathcal{V}^\dagger $ for some unitary frame transformation $\mathcal{V}$ perturbatively close to identity, and effective Hamiltonian $D$ that commutes with $\mathcal{X}_{\tau/2}=e^{i\pi N}$, i.e. has emergent Rydberg number parity symmetry.
Indeed, at the $\theta=\gamma/2$ point, we obtain
\begin{equation}
\begin{split}
    H_F &\approx \mathcal{V} D \mathcal{V}^\dagger \\
    \mathcal{V} &\approx e^{-iA^{(0)}} \\
    A^{(0)} &= -\frac{\epsilon}{4} \left(\tilde{N}\left(\frac{\tau}{4}\right) - \tilde{N}\left(-\frac{\tau}{4}\right)\right)\\
    \tau D^{(0)} &= - \gamma \tilde{N}(0) - \epsilon \left( \tilde{N}\left(\frac{\tau}{4}\right) + \tilde{N}\left(-\frac{\tau}{4}\right) \right)\\
    2i\tau D^{(1)} &= \epsilon\frac{\gamma}{2}  \left[ \tilde{N}(0), \tilde{N}\left(\frac{\tau}{4}\right) + \tilde{N}\left(-\frac{\tau}{4}\right) \right],
\end{split}
\end{equation}
where the symmetrization $\tilde{N}\left( \frac{\tau}{4} \right) + \tilde{N}\left(- \frac{\tau}{4} \right)$ ensures that $D$ has parity symmetry at both zeroth and first orders (as well as higher orders not computed here). $\mathcal{V}$ implements corrections to this effective Hamiltonian beyond the zeroth order, thereby recovering the leading order contributions to $H_F$ in the original frame, computed in \eqs{eqn:FM_expansion_0}{eqn:FM_expansion_1}. 

In a regime of small Floquet periods $\Omega\tau / 4 \ll 1$, we may perform a perturbative expansion of the time-evolved $\tilde{N}(t)$ operators, 
\begin{equation}
\begin{split}
    \tilde{N}(t) &= N - \frac{\Omega t}{2} \Hpyp + \frac{(\Omega t)^2}{4}  \left(\Hpzp - \Hpxyp \right) \\
    &\;\;\;\;+ \mathcal{O}\left((\Omega \tau)^3) \right),
\end{split}
\end{equation}
where $\Hpyp \equiv\sum_i P_{i-1} \sigma_i^y P_{i+1}$ enacts blockaded local spin-flips with phase, $\Hpxyp \equiv \frac{1}{2}\sum_i P_{i-1} \left( \sigma_i^x \sigma_{i+1}^x + \sigma_i^y \sigma_{i+1}^y \right) P_{i+2}$ is the blockaded nearest-neighbor spin-exchange interaction, and $\Hpzp \equiv \sum_i P_{i-1} \sigma_i^z P_{i+1}$ is a diagonal term that can be decomposed into the total Rydberg number operator and next-nearest neighbor ground-ground and Rydberg-Rydberg repulsion: $\Hpzp = - 3 N + \frac{1}{4} H_{ZIZ} + \text{const.}$, with $H_{ZIZ} = \sum_i \sigma^z_i \sigma^z_{i+2}$.
Inserting into \eqs{eqn:FM_expansion_0}{eqn:FM_expansion_1}, we thus obtain a closed-form expression of the effective Floquet Hamiltonian:
\begin{equation}\label{eqn:effective_model_app}
\begin{split}
    H_F &\approx -J\; N - h\; \Hpxyp + g\; \Hpxp + \frac{h}{4}\; H_{ZIZ} \\
    J &= \frac{(\gamma+2\epsilon)}{\tau} - \frac{3\epsilon \Omega^2 \tau}{32},\; h = - \frac{\epsilon \Omega^2 \tau}{32},\; g = - \frac{\epsilon(\theta+\epsilon)\Omega}{8},
   \end{split}\end{equation}
by keeping terms up to quadratic order in $\Omega\tau,\epsilon,\gamma,\theta$.

\section{Blockaded spin-exchange model with approximate $U(1)$ symmetry}

\subsection{Ground state phase diagram}
\label{sec:effective_model_phases}

The effective Hamiltonian $H_F$ described in \eq{eqn:effective_model_app} features total Rydberg number conservation at $g=0$, and is known to be integrable at this point~\cite{alcaraz1999}.
Furthermore, Ref.~\cite{verresen2019} explored this $U(1)$ symmetric model in the absence of the $H_{ZIZ}$-term: As a function of the chemical potential $J$, the model exhibits gapped paramagnetic and Néel-ordered phases, as well as an intermediate gapless Luttinger liquid phase. The Luttinger liquid is robust to $U(1)$-breaking single spin fluctuations when the Luttinger parameter $K<1/8$, and is stabilized by $\neel$ parity symmetry for $1/8<K<1/2$. For $K>1/2$, the Luttinger liquid is unstable towards $U(1)$-breaking perturbations even in the presence of $\neel$ symmetry and flows to a gapped topological phase.

Here, we adapt a similar analysis for $H_F|_{g=0}$ which additionally includes the $H_{ZIZ}$ term. Importantly, this term maintains integrability: Our system corresponds to a class of constrained XXZ models that Ref.~\cite{alcaraz1999} provides exact Bethe ansatz solutions for (concretely, $t=1$ and $\Delta=-0.5$ in~\cite{alcaraz1999}), which we review here for completeness. 
As the total Rydberg number $N \in \{0,\dots,L/2\}$ is a conserved quantity, each $N$-particle sector of the model can be diagonalized separately, by eigenstates labeled by quasi-momenta $\{k\}=k_1,k_2,\dots,k_N$ satisfying the following Bethe ansatz equations~\cite{alcaraz1999,bethe_ansatz_I_1997}:
\begin{equation}
    e^{ik_jL}=(-1)^{N-1} \prod_{l=1}^N e^{i (k_j-k_l)} \frac{1+ e^{ik_j}+e^{i(k_j+k_l)}}{1+ e^{ik_l}+e^{i(k_j+k_l)}},
\end{equation}
and with the following total energy and momentum:
\begin{align}
    E_k &= -2h \sum_{i=1}^N \cos{k_i}+\frac{h}{4}(L-4N) \\ 
    P_k&=\sum_{i=1}^N k_i.
\end{align}
Low-energy properties of this model can be studied in the thermodynamic limit, through integral equations for quasi-momenta distributions specified by the Bethe equations. In particular, the Luttinger parameter $K$ as a function of Rydberg density $n_0=N/L$ in the ground state reads:
\begin{equation}
\label{eqn:luttinger_param}
    K(n_0) = (1-n_0)^{2} \eta^{2}(U_0),
\end{equation}
where $\eta(U)$ and $U_0$ are determined by integral equations in Ref.~\cite{alcaraz1999}: 
\begin{align}
\begin{split}
Q(U) & =\frac{1}{2 \pi} \frac{\sin \pi/3}{\cos U-\cos \pi/3} \\
&\;\;\;\;-\frac{1}{2 \pi} \int_{-U_0}^{U_0} \frac{\sin (2 \pi/3) Q\left(U^{\prime}\right)}{\cosh \left(U-U^{\prime}\right)-\cos (2 \pi/3)} d U^{\prime} 
\end{split}\\
1 & =\eta(U)+\frac{1}{2 \pi} \int_{-U_0}^{U_0} \frac{\sin (2 \pi/3) \eta\left(U^{\prime}\right)}{\cosh \left(U-U^{\prime}\right)-\cos (2 \pi/3)} d U^{\prime},
\end{align}
which are subject to the constraint
\begin{equation}
\int_{-U_0}^{U_0} Q(U') dU'= \begin{cases}\frac{n_0}{1- n_0}, & 0 \leq n_0 \leq \frac{1}{2+n_0} \\ \frac{1-2 n_0}{1- n_0}, & \frac{1}{2+n_0} \leq n_0 \leq \frac{1}{1+n_0}\end{cases}.
\label{eqn:U0_constraint}
\end{equation}

\begin{figure}
    \centering
    \includegraphics[width=\linewidth]{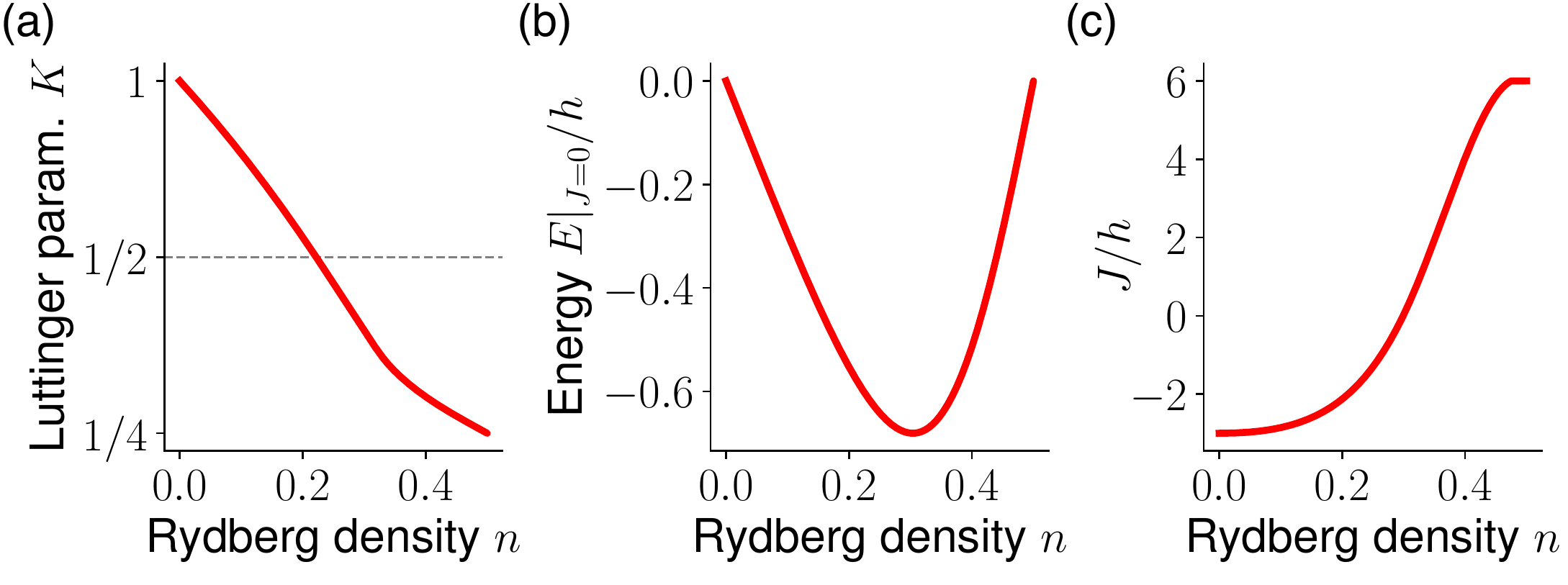}
    \caption{\textbf{Extracting the Luttinger parameter.} \textbf{(a,b,c)} The Luttinger parameter $K$ (\eq{eqn:luttinger_param}), the ground state energy $E|_{J=0}$ (\eq{eqn:ground_state_energy}) in the absence of chemical potential, and the chemical potential $J$ (\eq{eqn:chemical_potential}) as a function of ground state Rydberg density $n_0$. These quantities are obtained from the numerical solution of the integral equations in \eq{eqn:effective_model_app}.}
    \label{fig:s1}
\end{figure}

Moreover, the ground state Rydberg density $n_0$ at a given chemical potential $J$ is determined by minimizing the ground state energy as a function of Rydberg density. The ground state energy $E$ for our model is related to that of the unconstrained XXZ chain, $\tilde{E}$, provided in Ref.~\cite{Franchini_2017}:
\begin{align}
\label{eqn:ground_state_energy}
    E(n_0) &= (1-n_0) \tilde{E}(n_0) 
    \\
    \tilde{E}(n_0) &= L \int_{-U_0}^{U_0} \left(\epsilon(U') - J\right) Q\left(U^{\prime}\right) d U^{\prime}, \\
    \epsilon(U) &= -2h\frac{\sin^2 (\pi/3) } {\cosh(U) - \cos(\pi/3)},
\end{align}
where the factor of $(1-n_0)$ accounts for the reduced effective length of the chain due to the blockade constraint. This energy is minimized when 
\begin{align}
    \frac{\partial E}{\partial n_0}=-\tilde{E}(n_0) + (1-n_0) \frac{\partial \tilde{E}}{\partial U_0} \frac{\partial U_0}{\partial n_0} = 0.
\end{align}
Employing Leibniz integral rule on \eq{eqn:U0_constraint} gives
\begin{equation}
2 \frac{\partial U_0}{\partial n_0} Q(U_0) = \begin{cases}  \frac{1}{(1-n_0)^2}, & 0 \leq n_0 \leq \frac{1}{2+n_0} \\ \frac{-1}{(1- n_0)^2}, & \frac{1}{2+n_0} \leq n_0 \leq \frac{1}{1+n_0}\end{cases}.
\end{equation}
Furthermore, Ref.~\cite{Franchini_2017} derives
\begin{align}
    \frac{\partial \tilde{E}}{\partial U_0}=2L Q(U_0) \left[ -J \eta(U_0)-4\pi h \sqrt{1-\Delta^2} Q(U_0)\right].
\end{align}
Combining these, we arrive at the ground state condition
\begin{align}
\begin{split}
    \frac{\tilde{E}(n_0)}{L} &= \left( -J \eta(U_0)-4\pi h \sqrt{1-\Delta^2} Q(U_0)\right) \times \\
    &\;\;\;\;\; \times \begin{cases} \frac{+1}{1-n_0}, & 0 \leq n_0 \leq \frac{1}{2+n_0} \\ \frac{-1}{1-n_0}, & \frac{1}{2+n_0} \leq n_0 \leq \frac{1}{1+n_0}\end{cases}
\end{split}
     \\
    &= \frac{E|_{J=0}}{1-n_0} - J \times \begin{cases}\frac{n_0}{1- n_0}, & 0 \leq n_0 \leq \frac{1}{2+n_0} \\ \frac{1-2 n_0}{1- n_0}, & \frac{1}{2+n_0} \leq n_0 \leq \frac{1}{1+n_0}\end{cases},
\end{align}
where we define $E|_{J=0} \equiv (1-n_0)\int_{-U_0}^{U_0} \epsilon(U') Q\left(U^{\prime}\right) d U^{\prime}$, and thus
establish the relation between chemical potential $J$ and ground state Rydberg density $n_0$, as desired:
\begin{align}
    \label{eqn:chemical_potential}
    J &= \begin{cases} \frac{E|_{J=0} + 4\pi h \sqrt{1-\Delta^2} Q(U_0)}{n_0 - \eta(U_0)}, & 0 \leq n_0 \leq \frac{1}{2+n_0} \\ \frac{E|_{J=0} -4\pi h \sqrt{1-\Delta^2} Q(U_0)}{(1-2n_0)+ \eta(U_0)}, & \frac{1}{2+n_0} \leq n_0 \leq \frac{1}{1+n_0}\end{cases}.
\end{align}

By numerically integrating these equations using the method of quadratures, as shown in \fig{fig:s1}, we obtain the Luttinger parameter $K$ as a function of chemical potential $J$ in \figc{fig:3}{c} of the main text. Along with the ground state Rydberg density $n_0$, this enables characterizing the ground state phases of $H_F|_{g=0}$, sweeping $J/h$: The model exhibits a gapped $\neel$ phase with $n_0=1/2$ for $J/h>6$ and a gapped paramagnet with $n_0=0$ for $J/h<-3$. Between $-3<J/h<6$, where $0<n_0<1/2$, a gapless Luttinger liquid emerges. 
Following the arguments of Ref.~\cite{verresen2019}, the Luttinger liquid is stable to general $U(1)$ symmetry breaking perturbations for $K<1/8$ and stable to $U(1)$ symmetry breaking perturbations that preserve $\neel$ parity symmetry for $1/8<K<1/2$. For $K>1/2$, the Luttinger liquid is unstable towards such perturbations and flows to the gapped topological phase of the Kitaev chain~\cite{Kitaev_2001}.

\subsection{Domain wall dynamics from $\neel$ state}
\label{sec:domain_wall_dynamics}
In the main text, we considered quantum quenches of the $\neel$ state into the Luttinger liquid, enabled by turning on a small $U(1)$ symmetry breaking field $g$ that couples the $\neel$ state to other particle number sectors. This small field generates a low density of domain walls on top of the $\neel$ state, and majority of dynamics can be studied in the $0$-domain wall (N\'eel) and 2-domain wall sectors at sufficiently early times.

In order to study dynamics in this regime, we switch to a new description of $H_F|_{g=0}$ in \eq{eqn:effective_model_app} in terms of bond variables, with domain walls on top of a $\neel$ vacuum defined as the particle degrees of freedom. One subtlety of working with domain walls is to track the global $\neel$ gauge degree of freedom: There are two distinct vacuum states $|\neel\rangle$ and $|\neelp\rangle$, leading to two types of domain walls: $\neel \to \neelp$ domain walls on even bonds and $\neelp \to \neel$ domain walls on odd bonds, which cannot be converted into one another under blockade-consistent hopping. As such, each domain wall can be labeled by its location $d=0,\dots,L-1$, which contains two pieces of information $(u,\mu)$: the 2-site unit cell $u=\lfloor d/2 \rfloor=0,\dots,L/2-1$ it belongs to, and whether it is an even or odd-type domain wall, $\mu=e,o$.
We first compute the spectrum of the blockade-constrained interacting hopping model $H_F|_{g=0, J=0} = -h\Hpxyp + \frac{h}{4}H_{ZIZ}$. The zero-particle sector consists of the two vacuum states $|\neel\rangle$ and $|\neelp\rangle$ with energy
\begin{equation}
    \tilde{E}_{\varnothing} = \frac{h}{4}L.
\end{equation}
In the single-particle (single domain wall) sector, the Hamiltonian acts as 
\begin{equation}
    \begin{split}
    H_F|_{g=0,J=0}|u,\mu\rangle &= -h \left( |u-1,\mu\rangle + |u+1,\mu\rangle \right) \\
    &\;\;\;\;+ \frac{h}{4} (L-4) |u,\mu\rangle,
    \end{split}
\end{equation}
and can be exactly diagonalized using a plane wave ansatz
\begin{equation}
    |k,\mu \rangle = \sum_{u=0}^{L/2-1} e^{-iku} |u,\mu\rangle.
\end{equation}
Accordingly, the single particle dispersion $\tilde{E}_k$ is given by
\begin{equation}
\begin{split}
\tilde{E}_{k} = -2h \cos{k} + \frac{h}{4} (L-4).
\end{split}
\end{equation}
To solve the two-particle sector, we use the Bethe ansatz
\begin{equation}
    \begin{split}
    |k;k^{\prime}\rangle &= \sum_{u \leq u^{\prime}} e^{-i \left( ku + k^{\prime}u^{\prime}\right)} |u,o;u^{\prime},e \rangle \\
    &\;\;\;\;\;\;\;\;\;+ S(k,k^{\prime})e^{-i \left( k^{\prime}u + ku^{\prime}\right)} |u,o;u^{\prime},e \rangle,
    \end{split}
\end{equation}
where we make use of the fact that domain walls come in even-odd pairs and their relative ordering is fixed.
On an infinite chain, the eigenenergies $\tilde{E}_{k,k^\prime}$ with respect to $H_F|_{g=0,J=0}$ can be determined from the stationary Schrödinger equation for $|u-u^{\prime}| \gg 1$, when the two domain walls are far apart. This results in
\begin{equation} \label{eq:2dwtilde}
\begin{split}
\tilde{E}_{k,k^{\prime}} &= -2h \left( \cos{k} + \cos{k^{\prime}} \right) + \frac{h}{4} (L-8).
\end{split}
\end{equation}
Once the chemical potential $J$ is incorporated into $H_F|_{g=0}$, the vacuum and two-domain wall state energies read:
\begin{align}
    \label{eqn:Z2_energy}
    E_{\varnothing} &= -J\frac{L}{2} + \frac{h}{4}L \\
     \label{eqn:2DW_energy}
    E_{k,k^{\prime}} &= -J\left(\frac{L}{2}-1\right) -2h \left(\cos{k} + \cos{k^{\prime}} \right) + \frac{h}{4} (L-8).
\end{align}

With $\tilde{E}_{k,k^\prime}$ given by \eq{eq:2dwtilde}, the scattering phase $S(k,k')$ can be determined by projecting the stationary Schrödinger equation onto the state $\ket{u,o;u,e}$, where two domain walls are located in the same unit cell $u$. Specifically,
\begin{widetext}
\begin{equation}
\begin{split}
    \tilde{E}_{k,k^{\prime}}\braket{u,o;u,e|k;k^{\prime}} &= \bra{u,o;u,e}H_F|_{g=0,J=0}|k;k^{\prime}\rangle, \\
    \tilde{E}_{k,k^{\prime}} e^{-i\left(k+k^{\prime}\right)u} \left(1 + S(k,k^{\prime} )\right) &= \frac{h}{4} (L-4) e^{-i\left(k+k^{\prime}\right)u} \left(1 + S(k,k^{\prime} )\right) - \\
    -h \biggl( &e^{-i\left(k(u-1)+k^{\prime}u\right)} + S(k,k^{\prime} ) e^{-i\left(k^{\prime}(u-1)+ku\right)} + e^{-i\left(ku+k^{\prime}(u+1)\right)} + S(k,k^{\prime} ) e^{-i\left(k^{\prime}u+k(u+1)\right)} \biggr) \\
    \Rightarrow S(k,k^{\prime}) &= - \frac{e^{-ik} + e^{ik^{\prime}} + 1 }{ e^{ik} + e^{-ik^{\prime}} + 1 }.
\end{split}
\end{equation}
\end{widetext}

Equipped with the two-particle eigenstates $\ket{k;k^\prime}$, we introduce a weak $U(1)$ symmetry breaking perturbation $g\Hpxp$ and investigate the resulting dynamics starting from the $\ket{\neel}$ product initial state. In particular, the PXP perturbation connects $\ket{\neel}$ to the two-particle sector via
\begin{align}
    \Hpxp |\neel\rangle &= \sum_u |u,o,u,e\rangle.
\end{align}
Moreover, as both the Hamiltonian and initial state are two-site translationally invariant, the only non-vanishing couplings are to states $\ket{k;k^\prime}$ with $k^\prime = -k$, which exhibit the same translational invariance.
In particular, we consider the relevant matrix elements between $|\neel \rangle$ and the normalized two-domain-wall eigenstates $\frac{\mathcal{N}_{\mathrm{bc}}}{L}|k;-k\rangle$, where $\mathcal{N}_{\mathrm{bc}}$ is an $\mathcal{O}(1)$ normalization constant that depends on the boundary conditions; $\mathcal{N}_{\mathrm{obc}}=2$, $\mathcal{N}_{\mathrm{pbc}}=\sqrt{2}$.
They are given by
\begin{equation} \label{eqn:coupling_strength}
\begin{split}
 \lambda(k) \equiv   &\frac{\mathcal{N}_{\mathrm{bc}}}{L}\langle \neel  | H_F | k;-k \rangle = g\frac{\mathcal{N}_{\mathrm{bc}}}{L} \sum_{u} \bigl(  1+ S(k,-k) \bigr) = \\
    = & g \frac{\mathcal{N}_{\mathrm{bc}}}{2}\left( 1 - \frac{2e^{-ik} + 1 }{ 2e^{ik} + 1 }  \right) = g \frac{\mathcal{N}_{\mathrm{bc}}}{2} \frac{4i\sin{k}}{2e^{ik}+1},
\end{split}
\end{equation}
and we note that $\lambda(k)$ is proportional to the group velocity $2h\sin(k)$ of the single domain wall dispersion.
We thus see that the dynamics from $\ket{\neel}$ probes the low-energy spectrum of $H_F$ and couples to the two-domain-wall band at strength $\lambda(k)$ and energy offset $\delta(k) \equiv E_{k,-k} - E_{\varnothing} = J-2h-4h\cos(k)$.
When $\delta(k) \lesssim \lambda(k)$, domain walls are created slowly but resonantly, generating coherence between the two different $\neel$ orders and leading to the growing Quantum Fisher Information observed in \figc{fig:3}{d} of the main text.

\section{Dynamical probe of gapless phase}
\label{sec:dynamical_probe_gapless}

\begin{figure}
    \centering
    \includegraphics[width=\linewidth]{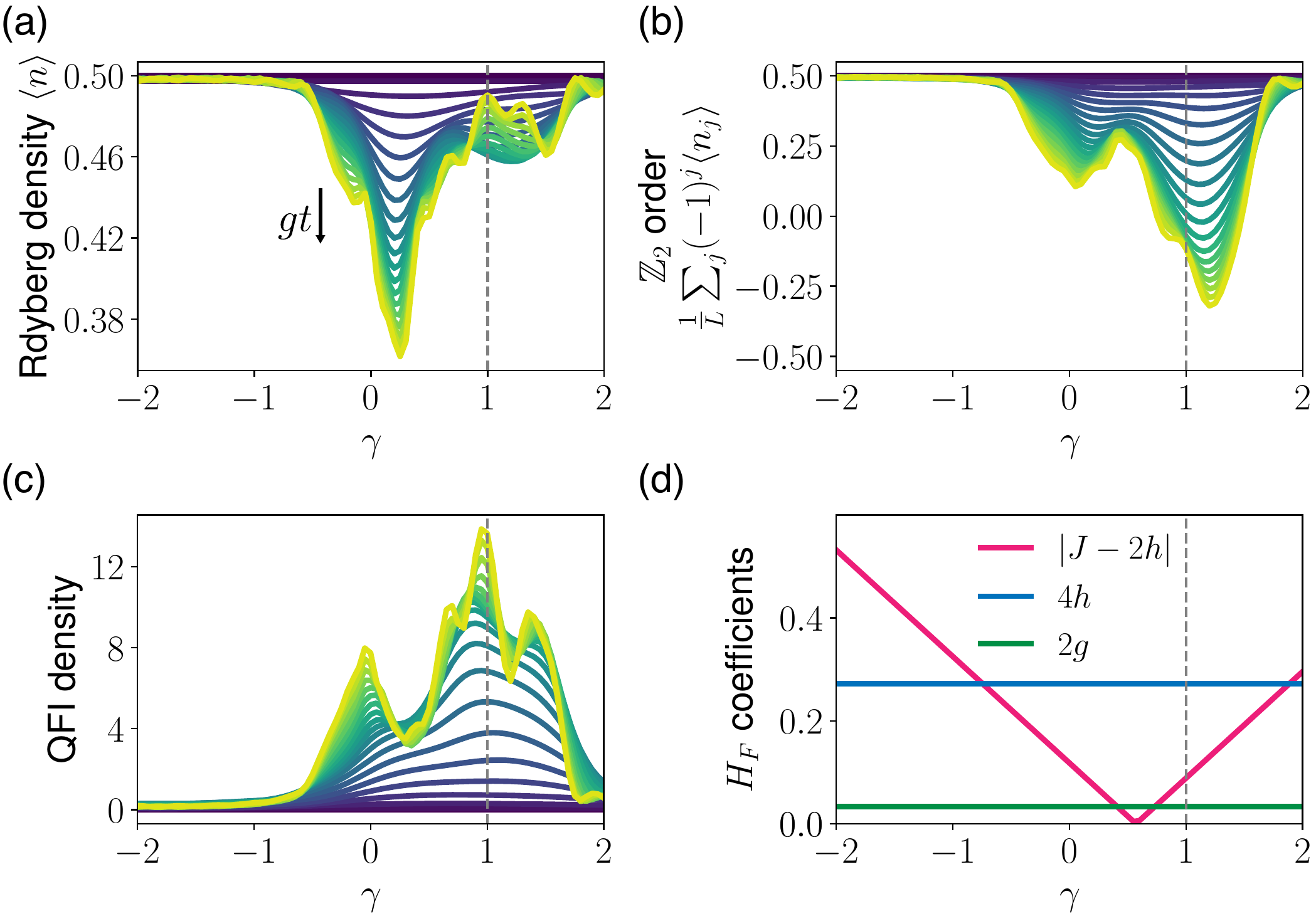}
    \caption{\textbf{Dynamical probe of gapless phase.} \textbf{(a,b,c)} Stroboscopic evolution of a $L=16$ periodic chain under a drive with period $\tau=2\pi/1.3$ and perturbations set to $\epsilon=-0.45$, $\theta=0.15$ while varying $\gamma$, which effectively sweeps $J$ for fixed $h,g$ in the effective Hamiltonian \eq{eqn:effective_model_app}. Within a suitable range of $\gamma$ values, dynamics starting from a N\'eel state generates domain walls that wash out the initial $\neel$ order and lead to a build-up of large Quantum Fisher information density. \textbf{(d)} Effective Hamiltonian coefficients $J,h,g$ evaluated for these drive parameters as a function of $\gamma$. The $\neel$ state is resonant with a segment of the two-domain wall band within $|J-2h|<|4h|$, which indeed aligns with the range of $\gamma$ values exhibiting growth of large multipartite entanglement.}
    \label{fig:s2}
\end{figure}

As shown in the previous section and verified in \fig{fig:3} of the main text, the creation of multipartite entanglement in the dynamics of the effective Hamiltonian $H_F$ is due to a weak but resonant coupling between the initial $\neel$ state and a low-energy two-domain-wall band. As such, the entanglement dynamics acts as a probe for the transition between a gapped $\neel$ state and a gapless Luttinger liquid of domain walls in $H_F$. 

Here, we verify numerically that these features are indeed also present in the corresponding stroboscopic dynamics of the full Floquet time evolution, even for large perturbations around the many-body echo.
Specifically, we consider the Floquet protocol with fixed parameters $\tau=2\pi/1.3$, $\epsilon=-0.45$, $\theta=0.15$, and a varying parameter $\gamma$ that controls the effective detuning/chemical potential in $H_F$ according to \eq{eqn:effective_model_app}.
Starting from $\ket{\neel}$, we consider the stroboscopic dynamics of the Rydberg density $\braket{n(t)}$, the $\neel$ order parameter $\frac{1}{L}\sum_j (-1)^j \braket{n_j(t)}$, and the QFI density in \fig{fig:s2}. 
Upon varying $\gamma$, we indeed find resonant excitation of domain walls and build-up of large multipartite entanglement within the regime corresponding to the Luttinger liquid phase of $H_F|_{g=0}$, see \figc{fig:s2}{d}. 

\begin{figure}
    \centering
    \includegraphics[width=\linewidth]{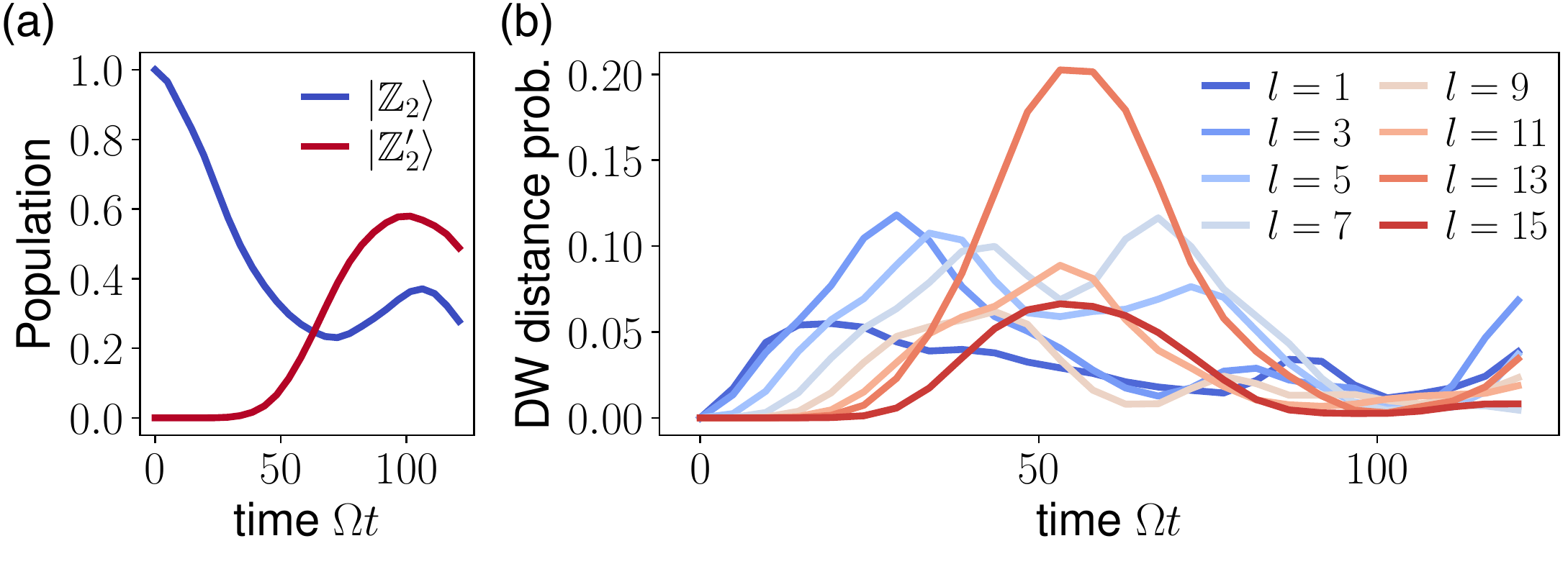}
    \caption{\textbf{Populations \& domain distance probabilities.} \textbf{(a)} $|\neel\rangle$ and $|\neelp\rangle$ populations during stroboscopic evolution of a $\neel$ state on $L=16$ periodic chain, under a Floquet drive with period $\tau=2\pi/1.3$ and perturbations set to $\epsilon=-0.45$, $\gamma=1.0, \theta=0.15$. \textbf{(b)} Probability of finding a pair of domain walls of type $(\neel\to\neelp,\neelp\to\neel)$ separated by a distance $l$, conditioned on the presence of at least one domain wall pair of this type.}
    \label{fig:s3}
\end{figure}

In order to highlight the role of domain wall pairs in mediating these entanglement dynamics, we study the stroboscopic time populations of $|\neel\rangle$, $|\neelp\rangle$, and two-domain wall states. In particular, we define a ``domain-wall distance probability'' for finding a pair of domain walls of type $(\neel\to\neelp,\neelp\to\neel)$ separated by a distance $l$, conditioned on the presence of at least one domain wall pair of this type. We note that this distribution is accessible in $z$-basis measurements, and directly captures large-scale fluctuations in the system for large $l$. 
\fig{fig:s3} illustrates that dynamics from the $\neel$ state initially generates domain wall pairs with small distance $l$, but rapidly evolves into a superposition of many different distances.
Finally, domain wall pairs that have reached $l=L-1$ on the periodic chain re-annihilate to form a $\neelp$ state, in macroscopic superposition with the initial $\neel$ state, thus generating a GHZ state.

We note that in order to detect and quantify multipartite entanglement in an experimental setting, one also needs to measure off-diagonal observables probing the coherence of the open system, which can be achieved using a combination of native quenches~\cite{omran2019} and learning techniques~\cite{huang2020,elben2023,notarnicola2023,hu2024}.

\section{Performance of Floquet Protocol}
\label{sec:performance_floquet}
\begin{figure}[t]
    \centering
    \includegraphics[width=\linewidth]{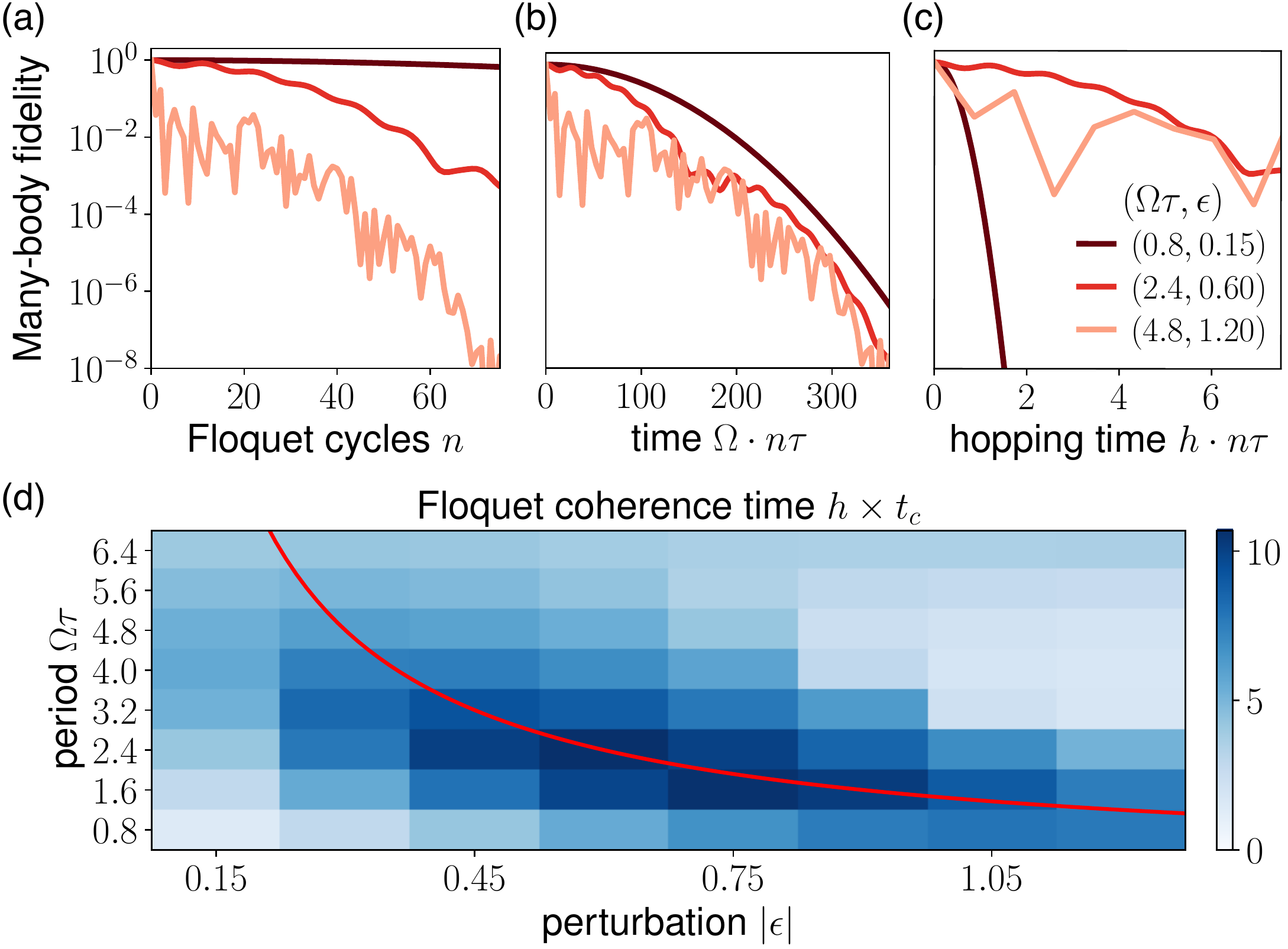}
    \caption{\textbf{Robustness and efficiency of Floquet protocol.} \textbf{(a,b,c)} Many-body state fidelities of \eq{eqn:floquet_coherence_app} evaluated for stroboscopic evolution under Floquet protocols that realize a single-particle quantum walk on an $L=16$ periodic chain. The drive is given by $-2\epsilon=\gamma=2\theta$, generating an effective Hamiltonian corresponding to a pure hopping model with $h=-\frac{\epsilon\Omega^2\tau}{32}$ and $J=g=0$, see \eq{eqn:effective_model_app}.
The decay curves, which incorporate a phenomenological decay $t_*=100 \Omega^{-1}$, are presented as a function of (a) number of Floquet cycles, (b) physical time in units of the Rabi frequency, and (c) effective time in units of the hopping strength $h$ of the effective Hamiltonian. The results demonstrate the tradeoff between robust drives with small period $\tau$ and perturbation $\epsilon$, and efficient drives with larger perturbations that generate significant many-body dynamics within shorter amounts of physical time. Our definition for the effective coherence time 
incorporates both aspects.
\textbf{(d)} Coherence time $t_{c}$ (see \eq{eqn:floquet_coherence_app}) of the Floquet protocol as a function of drive period $\tau$ and perturbation strength $|\epsilon|$ for the drive parameterization considered in (a-c).
    We estimate an optimal $h\times t_{c} \approx 10$ in units of $h$.
    Red curve depicts a contour of constant $h$. 
}
    \label{fig:s4}
\end{figure}

We benchmark the Floquet protocol by evaluating the agreement between stroboscopic Floquet evolution $\ket{\Psi(n\tau)}$ and target effective Hamiltonian evolution $\ket{\Psi_F(n\tau)}$ for the quantum walk of a single Rydberg excitation. This is achieved by choosing the drive parameters as $-2\epsilon=\gamma=-\theta$, which generates a pure spin exchange model with $h=-\frac{\epsilon\Omega^2\tau}{32}$ and $J=g=0$ in \eqref{eqn:effective_model_app}.

Using state fidelity as a metric, we extract an effective coherence time $t_{c}$ from the decay
\begin{equation}
    |\langle \Psi_{F}(n\tau)|\Psi(n\tau)\rangle|^2 e^{-L(n\tau/t_*)^2} \sim e^{-L( n\tau/t_{c})^2},
\label{eqn:floquet_coherence_app}
\end{equation}
which incorporates a phenomenological decay constant $t_*=15 (2\pi/\Omega)$ to model finite coherence time of an experimental device. 
(The choice of an overall Gaussian decay is motivated by the numerically evaluated fidelities displayed in \figc{fig:s4}{a-c}.) Then, we vary the pulse parameters $\tau$ and $|\epsilon|$ to maximize $h\times t_{c}$, which measures how much hopping occurs before the system decoheres (see \figc{fig:s4}{d}). 

Here, we highlight the tradeoff between robustness and efficiency inherent to our Floquet protocol: 
On one hand, detuning profiles with small drive period $\tau$ and small perturbation $|\epsilon|$ result in a high fidelity between the stroboscopic dynamics and the evolution under $H_F$. On the other hand, the corresponding hopping $h \sim \epsilon \tau$ in $H_F$ is small, thus leading to slow dynamics that is eventually limited by physical coherence times.
To demonstrate this tradeoff, we show the decay of the many-body fidelity of \eq{eqn:floquet_coherence_app} for varying drive parameters and different choices of units of time. As expected, smaller perturbations $\epsilon$ and Floquet periods $\tau$ result in a higher fidelity after a given number of Floquet cycles $n$, see \figc{fig:s4}{a}. However, when plotted against physical time in units of the Rabi frequency $\Omega\tau n$, the relative fidelities for drives with different periods $\tau$ are altered significantly, see \figc{fig:s4}{b}. 
Finally, identifying the time $h \times n\tau$ in units of the hopping strength $h$ of the effective Hamiltonian $H_F$ as the most relevant scale for the quantum simulation of $H_F$, we see in \figc{fig:s4}{c} that Floquet protocols with sizeable parameters $\epsilon$, $\tau$ can outperform protocols with very weak perturbations around the many-body echo. Consequently, the Floquet coherence time $h\times t_{c}$ extracted from \eq{eqn:floquet_coherence_app} exhibits a non-monotonic dependence on pulse parameters $|\epsilon|$ and $\tau$, as shown in \figc{fig:s4}{d}, indicating an optimal choice of drive profile for the relevant experimental timescales.

\clearpage

\end{document}